\documentclass[11pt]{article}

\usepackage[dvips]{graphicx}
\usepackage{amsfonts}

\newcommand{\staccrel}[2]{\mathrel{\mathop{#1}\limits_{#2}}}
\newcommand{\beq}{\begin{equation}}
\newcommand{\eeq}{\end{equation}}
\newcommand{\Real}{\mbox{Re\,}}
\newcommand{\Imag}{\mbox{Im\,}}
\newcommand{\Dt}{\mbox{det\,}}
\newcommand{\Sg}{\mbox{sgn\,}}
\newcommand{\Cs}{\mbox{const.\,}}
\newcommand{\R}{{\mathbb{R}}}
\newcommand{\N}{{\mathbb{N}}}
\newcommand{\bx}{{\bf x}}
\newcommand{\rmi}{{\rm i}}
\newcommand{\rmd}{{\rm d}}
\newcommand{\SCR}{\parbox{1.09cm}{\scriptsize \baselineskip 8pt after the crossing} }

\newsymbol\square 1003

\begin{document}

\title{\bf{Orbiting Resonances and Bound States in Molecular Scattering}}

\author{\vspace{5pt} Enrico De Micheli$^{\dag}$ and Giovanni Alberto Viano$^{\ddag}$ \\
$^{\dag}$\small{Istituto di Cibernetica e Biofisica - Consiglio Nazionale delle Ricerche}\\[-5pt]
\small{Via De Marini, 6 - 16149 Genova, Italy. \vspace{8pt}} \\
$^{\ddag}$\small{Dipartimento di Fisica - Universit\`a di Genova} \\[-5pt]
\small{Istituto Nazionale di Fisica Nucleare - sez. di Genova} \\[-5pt]
\small{Via Dodecaneso, 33 - 16146 Genova, Italy.}
}

\date{January 11, 2002}

\maketitle

\begin{abstract}
A family of orbiting resonances in molecular scattering is globally described by using a
single pole moving in the complex angular momentum plane. The extrapolation of this pole at
negative energies gives the location of the bound states. Then a single pole trajectory,
that connects a rotational band of bound states and orbiting resonances, is obtained. These complex angular
momentum singularities are derived through a geometrical theory of the orbiting. The downward
crossing of the phase--shifts through $\pi/2$, due to the repulsive region of the molecular potential,
is estimated by using a simple hard--core model. Some remarks about the difference between diffracted rays 
and orbiting are also given.
\end{abstract}


\section{Introduction}
\label{introduction_section}
After a long search, about two decades ago it became possible to observe orbiting resonances (or quasi--bound
states) in molecular beam scattering \cite{Bernstein2,Weach,Schutte1,Schutte2,Toennies}. 
Although these phenomena can be quite naturally interpreted in the framework of scattering theory \cite{Child}, 
their semiclassical nature calls for a more refined analysis, mainly geometrical, which has been partially performed 
by several authors, and notably by Berry \cite{Berry1,Berry2,Berry3}. In this spirit, it has been also 
advocated the use of the complex angular momentum plane polology, especially by Bosanac 
\cite{Bosanac1,Bosanac2,Bosanac3}, Connor \cite{Connor1,Connor2} and Nussenzveig (see in particular his 
book \cite{Nussenzveig1} and the references quoted therein). In the latter approach, however, some 
problems remain, and a detailed phenomenological analysis is in part still missing.

In the standard complex angular momentum approach one generally considers a given class of potentials, and
then explores the analytical properties of the partial--waves along with their asymptotic behaviour in the complex
angular momentum plane. If these analytical properties and asymptotic behaviours allow for using the Watson 
transformation \cite{Watson}, then a resummation of the partial wave expansion can be performed, and the total 
scattering amplitude can be represented in terms of poles along with a background integral. 
The main limit of this method consists in the fact that
it works only for a rather restricted class of potentials, notably for the Yukawian class \cite{DeAlfaro},
which is not very important in molecular scattering. Nussenzveig \cite{Nussenzveig2,Nussenzveig3,Nussenzveig4}, 
in a series of very significant papers, extended the complex angular momentum method with particular 
attention to optical problems.
In this context, the hard--core potential has been treated in detail, and, in addition to the Watson resummation,
also the Poisson transformation has been used (see also Ref. \cite{Berry3}).
The latter seems to be more promising for semiclassical scattering theories like the one that is of 
interest in the present paper. However, it must be noted that
molecular scattering is usually described in terms of Lennard--Jones or Morse--type potential
\cite{Pauly}; it is worth reminding that singular potentials of the type $g^2r^{-n}$, $n>2$, present an 
infinite number of poles in the first quadrant of the complex angular momentum plane, and, in such a situation,
one is forced to use a modified Watson transformation. As far as we know, this approach has been rigorously 
proved in the case $g^2r^{-4}$ by Dombey and Jones \cite{Dombey}, and then conjectured for $g^2r^{-n}$, $n>4$. 
Furthermore, several authors, and notably those from the Uppsala school \cite{Thylwe1,Thylwe2,Froman}, 
have applied the phase--integral method to derive locations and residues of the poles in the complex 
angular momentum plane, for singular potentials of the form $V(r)=g^2 r^{-n} f(r)$, $(n \geq 4)$.
It should be noted that the primary goal of all these researches consists in substituting to the
partial--wave expansion, in a region where it is slowly convergent, more rapidly convergent representations.
The latter then split the scattering amplitude into a sum of pole contributions and one or more integral 
terms which are usually evaluated by means of the saddle point method.

Our viewpoint is quite different. We think that also in the region where only one partial--wave is dominant a 
complex angular momentum representation can be useful if we want to interpolate the various orbiting resonances
with a single moving pole, whose extrapolation at negative energies can then give the bound states. 
With this in mind, we focus our attention on the poles of the scattering amplitude in the complex angular 
momentum plane, and we neglect, to a certain extent, the background integral, as will be explained below. 
To this purpose, a geometrical method of derivation of the complex angular momentum poles
is implemented: instead of studying the analytical and asymptotic properties of the partial--waves 
for a specific class of potentials, we analyze the particle motion (and specifically the orbiting) in a 
Riemannian space whose metric is induced by the potential.

In view of the fact that we are treating a semiclassical scattering problem, the concept of particle path
(or trajectory) still conserves its meaning. We are therefore allowed to study the geodesics (i.e., the
particle paths) in a space whose Riemannian metric is given by $n^2\,dx^i\,dx^j$, $n^2=1-V/E$, $E$ being
the energy and $V$ the potential. The coordinates $x^i$ require a more detailed comment: in fact, one generally
claims to solve the dynamical problem in the whole space $\R^3$ by using a unique coordinate system
for all the points of the space being considered. Here we prefer to introduce local coordinates that are 
appropriate
to describe locally the particle trajectories, and then to use different coordinate systems elsewhere
in space. We are thus led to consider the Jacobian of the transformation connecting the local coordinates
to the ambient space coordinates. The transformation is singular in those domains where the Jacobian vanishes.
We can then work out the problem by means of two geometrical tools: geodesics, which describe locally the 
particle paths, and Jacobians. The global solution emerges by patching up geodesic segments which represent
local solutions. We will show in section \ref{outline_section} that by the sole use of these geometrical 
ingredients is possible to obtain a representation of the scattering amplitude in terms of complex angular 
momentum plane singularities (poles).

However, this representation is not complete, and can only be regarded as an approximation; in what follows we shall
discuss pro and con of this representation.
First of all we obtain an angular distribution in terms of Legendre functions (not polynomials) of complex index, 
that present a logarithmic singularity forwards (i.e., at null scattering angle). This fact suggests that a 
compensating background term is missing in a representation which uses only poles. However, it is possible to
project the scattering amplitude (represented only by poles) on the partial--waves, and the result 
can be regarded as a faithful representation, at least for angular momentum sufficiently small. 
Furthermore, the consequent integral (total) cross--section is finite, since the singularity of the Legendre 
functions involved is only logarithmic. 
On the other hand, the advantage of this representation is that the entire
sequence of orbiting resonances can be fitted by using a single pole.
The latter, while moving as a function of the energy, is able to describe in an ordered way the passage 
of the phase--shifts of various angular momenta through $\pi/2$, that is
the value of the phase--shifts corresponding to the resonances. Therefore, a global view and explanation 
of the orbiting resonances is obtained.

Furthermore, these moving poles produce bound--states at those values of the energy where the angular momentum 
(which is real at negative energy) takes integer values. We thus obtain a trajectory of the complex angular 
momentum pole 
as a function of the energy, which gives bound--states at negative energy and resonances at positive energy.
Obviously, the bound--states are not observed directly by scattering experiments. 
Their existence is, however, revealed by the fact that the phase--shift $\delta_l$ at zero--energy is given 
by $n_l\pi$ (as explained by the generalized Levinson theorem) where $n_l$ is the number of bound--states 
with angular momentum equal to $l$ (see refs. \cite{Levinson,Swan}, and sections \ref{orbiting_section} and 
\ref{phenomenological_section}). 
These non--resonating phase--shifts decrease at higher energy passing through $\pi/2$ in downward direction. 
This downward crossing of $\pi/2$ does not correspond to a resonance, but it is rather due to a repulsive potential.
We can, however, fit this part of the phase--shifts by regarding them as the ones produced by a repulsive hard--core 
in a sense that will be explained in detail in section \ref{orbiting_section}.

The paper is organized as follows: in section \ref{outline_section} we outline a geometric theory of the orbiting,
and obtain an approximate representation of the scattering amplitude in terms of moving poles (complex 
angular momentum poles). In section \ref{orbiting_section} we derive a fitting formula for the phase--shifts 
that accounts also for the role of the non--resonating phase--shifts. Section \ref{phenomenological_section} is 
devoted to the phenomenological analysis. Finally, in section \ref{remarks_section} we draw some conclusions, 
and briefly discuss the difference between orbiting and diffracted rays, and, accordingly, between resonances 
and diffraction.

\section{Outline of the geometrical theory of the orbiting}
\label{outline_section}
Let us start from the Helmholtz's equation
\beq
\label{uno}
\Delta u + k^2n^2 u = 0\,,~~~~~\left(n^2=1-\frac{V}{E}\right),
\eeq
where 
$E$ is the energy and $V$ the potential. Now, we look for a solution of equation (\ref{uno})
of the following form:
\beq
\label{due}
u(\bx,k) = \int A(\bx,\beta) e^{\rmi k\Phi(\bx,\beta)}\,\rmd\beta\,,~~~~~(\bx\in{\R} ^3).
\eeq
The principal contribution to $u(\bx,k)$, as $k\rightarrow\infty$, corresponds to the stationary points of $\Phi$,
in the neighborhoods of which the exponential $\exp(\rmi k\Phi)$ ceases to oscillate rapidly. These stationary points
can be obtained from the equation $\partial\Phi(\bx,\beta)/\partial\beta = 0$ (assuming that
$\partial^2\Phi(\bx,\beta)/\partial\beta^2 \neq 0$). Then, the following asymptotic expansion of $u$, as 
$k\rightarrow\infty$, is valid:
\beq
\label{tre}
u(\bx,k) \simeq e^{\rmi k\Phi(\bx,\beta_0)}\sum_{m=0}^\infty\frac{A_m}{(\rmi k)^m}\,,
\eeq
where $\beta_0$ is the unique stationary point of $\Phi$ at fixed $\bx$. The leading term of expansion (\ref{tre})
reads
\beq
\label{quattro}
u(\bx,k) = A_0(\bx) \, e^{\rmi k\Phi(\bx,\beta_0)},
\eeq
where
\beq
\label{cinque}
A_0(\bx) = A(\bx,\beta_0)\left(\left|\frac{\partial^2\Phi}{\partial\beta^2}\right|^{-1/2}\right)_{\beta=\beta_0}
\exp\left\{\rmi\frac{\pi}{4}\Sg\left(\frac{\partial^2\Phi}{\partial\beta^2}\right)_{\beta=\beta_0}\right\}.
\eeq
The leading term (that hereafter will be written as $Ae^{\rmi k\Phi}$, omitting the subscript zero) can be substituted 
into equation (\ref{uno}) and then, by collecting the powers of $(\rmi k)$ and nulling their coefficients, 
we obtain two equations: \\
a) the eikonal equation
\beq
\label{sei}
\left(\nabla\Phi\right)^2=n^2=1-\frac{V}{E}\,;
\eeq
b) the transport equation
\beq
\label{sette}
\nabla\cdot(A^2\nabla\Phi) = 0\,.
\eeq
The physical meaning of the transport equation (\ref{sette}) is that the probability current density
is conserved. In order to exploit the geometrical content of the eikonal equation, it is
convenient to replace the optical index $n^2$ with a Riemaniann metric tensor $g_{ij}$. Then, the eikonal
equation can be rewritten as follows:
\beq
\label{otto}
g^{ij}\frac{\partial\Phi}{\partial x^i} \frac{\partial\Phi}{\partial x^j} = 1\,,~~~~~
\left(g^{ij} = \left(g_{ij}\right)^{-1}\right),
\eeq
$\{x^i\}$ representing a system of local coordinates. Accordingly, the transport equation will be rewritten
in the following form:
\beq
\label{nove}
\frac{1}{\sqrt g}\sum_i \frac{\partial}{\partial x_i}\left\{\sqrt{g}\left(A^2
\sum_j g^{ij} \frac{\partial\Phi}{\partial x_j}\right)\right\} = 0\,,
\eeq
where $g=|\Dt(g_{ij})|$. Now, we consider the tube composed by the trajectories which describe only circular
orbits and assuming, at this stage, that there is no leakage from the tube and, accordingly, 
there is no attenuation in the ray tube intensity. 
The angular coordinates used are $\theta$ and $\phi$ and, for the sake of simplicity, we set
the radius of the orbit to $1$; then we have: $g_{11}=1, g_{22}=\sin^2\theta, g_{12}=g_{21}=0$.
From Eqs. (\ref{otto}) and (\ref{nove}) we get
\begin{eqnarray}
\label{dieci}
&&\left(\frac{d\Phi}{d\theta}\right)^2=1, \\
\label{undici}
&&\frac{1}{|\sin\theta|}\left\{\frac{d}{d\theta}\left(|\sin\theta|A^2\frac{d\Phi}{d\theta}\right)\right\}=0,
~~(\theta\neq n\pi),
\end{eqnarray}
where we supposed that both the phase $\Phi$ and the amplitude $A$ do not depend on $\phi$.

From equation (\ref{dieci}) we have: $\Phi=\pm\theta+\Cs$. Next, substituting equation (\ref{dieci}) into equation (\ref{undici})
we finally obtain (see formula (\ref{quattro}))
\beq
\label{dodici}
u(\theta,k)=\frac{\Cs}{\sqrt{|\sin\theta|}}\,e^{\pm\rmi k\theta}\,,~~~~~(\theta\neq n\pi, n=0,1,2,\ldots),
\eeq
where the terms $e^{\pm\rmi k\theta}$ represent waves traveling in counterclockwise sense ($e^{\rmi k\theta}$) or
in clockwise sense ($e^{-\rmi k\theta}$). At this point let us note that approximation (\ref{dodici}) fails
at $\theta=n\pi$ ($n=0,1,2,\ldots$): i.e., this approximation is not uniform. In particular, we are obliged
to find the connection formulae relating the traveling waves when they cross the antipodal points
$\theta=0$, and $\theta=\pi$. These connection formulae can be established by the use of the Maslov
indexes. Here we simply give the results (the interested reader is referred to refs. \cite{DeMicheli,Maslov}):
\begin{itemize}
\item[i)] for the counterclockwise orbiting wave we have, after the crossing through the point $\theta=0$,
\beq
\label{tredici}
e^{\rmi k\theta}~~ \stackrel{\SCR}{\displaystyle \put(0,0){\vector(1,0){40}}} ~~e^{\rmi(k\theta -\frac{\pi}{2})}\,;
\eeq
\item[ii)] for the clockwise orbiting wave we have, after the crossing through the point $\theta=0$,
\beq
\label{quattordici}
e^{-\rmi k\theta}~~ \stackrel{\SCR}{\displaystyle \put(0,0){\vector(1,0){40}}} ~~e^{-\rmi(k\theta -\frac{\pi}{2})}\,.
\eeq
\end{itemize}
Therefore, for each complete tour, both the counterclockwise and the clockwise oriented rays cross the
antipodal points $\theta=0$ and $\theta=\pi$, and accordingly, in both cases the amplitude acquires
a factor $(-1)$.

We have seen that at the antipodal points the amplitude becomes infinite if we use the $\theta$--representation.
But in the Maslov construction use is made of the $p_\theta$--representation ($p_\theta$ being the variable
conjugate to $\theta$), and in the $p_\theta$--representation the intensity of the ray--tube is 
finite \cite{Maslov}.

In the case of unstable states the probability current density is not conserved and a tunnelling through the 
centrifugal barrier can, indeed, produce an attenuation of the intensity of the ray tube. The phenomenon can
be represented by picturing trajectories which leave tangentially the circular orbit and emerge at large distances.
In this process the whole $\R^3$ space is involved and we must analyze the Jacobian $J$ of the transformation
relating the local coordinates, which are appropriate for describing the orbiting, to the coordinates 
of the ambient space, that is the whole $\R^3$ space equipped with the euclidean metric.
The leakage from the tube of circular trajectories can be described as follows. 
A beam of particles enter the interaction region, describe circular orbits of radius $R$, and, after a certain 
number of revolutions can emerge and leave tangentially the interaction region. 
In atomic scattering, the latter is not well--defined because of the
long tail of the potentials. However, we can regard as the domain of interest for the orbiting process, the
region delimited by the centrifugal barrier from one side, and by the hard--core (or
more precisely by the high order singularity of the potential at the origin) from the other side. 
It is precisely in this
domain that the orbiting resonances take place; the lifetime of these states is then related to the probability of
tunnelling across the centrifugal barrier. The ambient space is properly described by the cartesian coordinates
$(x,y,z)$, the $z$--axis being directed along the incident beam and positively oriented in the direction of the
outgoing particles. The local coordinates ($\theta_0,\phi_0,\tau$) are defined as follows:
$\theta_0$ is the angle measured along the meridian circle from the point of incidence of the ray; $\phi_0$
is the azimuthal angle, and $\tau$ is a parameter along the ray tangential to the circular orbit. 
We can then evaluate the Jacobian (see also Ref. \cite{DeMicheli}):
\beq
\label{quindici}
J=\frac{\partial(x,y,z)}{\partial(\theta_0,\phi_0,\tau)}=\tau (R\cos\theta_0-\tau\sin\theta_0).
\eeq
The domain where the Jacobian vanishes is composed by:
\begin{itemize}
\item[$\alpha$)] the surface $\tau=0$;
\item[$\beta$)] the semi--axis $(z\geq 0, x=0, y=0)$ represented by $\tau=\bar{\tau}=R\cot\theta_0$.
\end{itemize}
When the Jacobian vanishes the eikonal approximation fails (for a detailed discussion of this very delicate 
mathematical point see Ref. \cite{DeMicheli}); furthermore, on the surface $\tau=0$ there is an 
attenuation of the intensity flux. We shall return below on this point. On the other hand, we can still use
the transport equation written in the following form:
\beq
\label{sedici}
\frac{1}{J}\frac{d}{d \tau} (J A^2) = 0 \,,
\eeq
whenever $J \neq 0$, and the current density is conserved. In particular, in the asymptotic region we can write:
$A=\Cs/\sqrt{J}$.

\begin{figure}[ht]
\vspace{1in}
\centering
\includegraphics[width=6in]{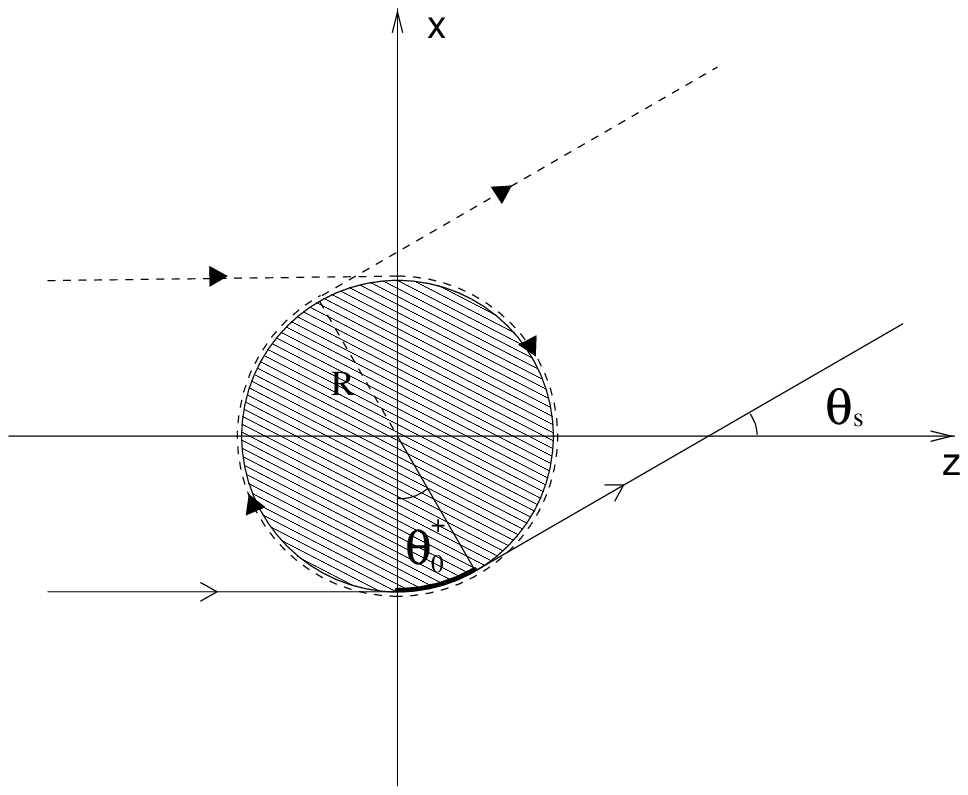}
\caption{
Geometrical model of orbiting:
the ray below (solid line with open arrows) travels $\theta_0^+$ radians in counterclockwise sense;
the ray above (dashed line with filled arrows) travels in clockwise sense and crosses the axial 
caustic (i.e., the $z$--axis) twice before emerging in the direction 
of the scattering angle $\theta_s$. The angle $\theta_0^+$ coincides with the scattering angle $\theta_s$.
}
\label{f1}
\end{figure}

Now, since $\tau=\sqrt{r^2-R^2}$, $r$ being the radial variable ($r^2=x^2+y^2+z^2$), then
$\tau$ tends to $r$ as $r\rightarrow\infty$. Accordingly, we obtain: $\sqrt{J}\rightarrow \rmi r\sqrt{\sin\theta_0}$ 
as $r\rightarrow\infty$. At this point, in order to evaluate the contribution to the wavefunction, at large
values of $r$, of the various classes of trajectories, we must distinguish between counterclockwise and
clockwise oriented orbits. We start considering the contribution of the counterclockwise trajectories, that 
leave the circular orbit without completing one tour, and that cross the $z$--axis only once. We have,
for large values of $r$,
\beq
\label{diciassette}
u_0(k,r,\theta_0^+) \staccrel{\sim}{r\rightarrow\infty}
a(k,\theta_0^+)\,\frac{e^{\rmi k(R\theta_0^+ +r)}}{\rmi r\sqrt{\sin\theta_0^+}}
= a(k,\theta_0^+)\, e^{-\rmi\pi/2}\, \frac{e^{\rmi kr}}{r} \frac{e^{\rmi kR\theta_0^+}}{\sqrt{\sin\theta_0^+}}\, ,
\eeq
$(0<\theta_0^+<\pi)$. Formula (\ref{diciassette}) requires some comments. Hereafter we shall use a superscript $(+)$
for all the quantities that refer to counterclockwise trajectories, and a superscript $(-)$ for all what
refers to clockwise oriented rays; the subscript $(0)$ in the notation of $u$ refers to the fact that $u_0$ accounts
only for the trajectories that have not completed one circular orbit. The factor $a(k,\theta_0^+)$ describes
the leakage of the particles by the tunnelling effect. The phase $\Phi$ (see formula (\ref{quattro})) now
reads: $\Phi(\theta_0^+,r) = R\theta_0^+ +\tau \staccrel{\sim}{r\rightarrow\infty} R\theta_0^+ + r$, and
the factor $e^{-\rmi\pi/2}$ can be regarded as the Maslov phase--shift due to the crossing of 
the trajectory across the $z$--axis,
where the Jacobian $J$ vanishes (see formula (\ref{tredici})). We can now observe that the angle $\theta_0^+$
coincides with the scattering angle $\theta_s$ (see figure \ref{f1}); we can then evaluate the contribution
to the scattering amplitude due to the counterclockwise trajectories which have not completed one orbit. We obtain,
with obvious notation:
\beq
\label{diciotto}
f_{(0)}^+(k,\theta_s)=a(k,\theta_s)\,
\frac{e^{-\rmi\pi/2}\,e^{\rmi kR\theta_s}}{\sqrt{\sin\theta_s}}\,,~~~~~(0 < \theta_s < \pi).
\eeq
Similarly, we can evaluate the contribution to the scattering amplitude of a clockwise trajectory which has not 
completed one tour. By observing that $\theta_0^-=2\pi-\theta_s$, we have
\beq
\label{diciannove}
f_{(0)}^-(k,\theta_s)=(-1)\,a(k,\theta_s)\,
\frac{e^{\rmi kR(2\pi-\theta_s)}}{\sqrt{\sin\theta_s}}\,,~~~~~(0 < \theta_s < \pi).
\eeq
Notice that the factor $(-1)$ is due to the product of two Maslov phase factors, in view of the fact that
the clockwise trajectory crossed the $z$--axis twice, even if it has not completed one orbit.

Adding $f_{(0)}^+$ to $f_{(0)}^-$, we obtain the contribution of all the rays that have not completed one orbit:
\beq
\label{venti}
f_{(0)}(k,\theta_s)=f_{(0)}^+ + f_{(0)}^- = 
-\rmi a(k,\theta_s)\, \frac{e^{\rmi kR\theta_s}-\rmi e^{\rmi kR(2\pi-\theta_s)}}{\sqrt{\sin\theta_s}},
~~~(0 < \theta_s < \pi).
\eeq
We must now take into account the contribution of all those trajectories that describe more than one complete 
circular orbit before emerging. Let us assume that they do $m$ ($m\in\N$) complete tours. The angles
$\theta_{0,m}^\pm$ are related to the scattering angle $\theta_s$ as follows:
\begin{eqnarray}
\label{ventuno}
\theta_{0,m}^+ & = & \theta_s + 2\pi m\,,~~~~~~(m=0,1,2,\ldots), \\
\label{ventidue} 
\theta_{0,m}^- & = & 2\pi - \theta_s + 2\pi m\,.
\end{eqnarray}
We have, for $0<\theta_s<\pi$,
\beq
\label{ventitre}
f(k,\theta_s)=-\rmi \sum_{m=0}^\infty (-1)^m\,e^{\rmi 2\pi mkR}\,a(k,\theta_s)\,
\frac{e^{\rmi kR\theta_s}-\rmi e^{\rmi kR(2\pi-\theta_s)}}{\sqrt{\sin\theta_s}}\,.
\eeq
Once again the factor $(-1)$, at each $m$, is due to the product of two Maslov phase--factors, 
corresponding to the fact that
both the counterclockwise and the clockwise rays cross the $z$--axis twice for each complete circular
orbit. The term $a(k,\theta_s)$ can be regarded as a damping factor that gives the attenuation of the tube 
of circular orbits due to the tunnelling through the centrifugal barrier. At fixed $k$ we can thus represent
the leakage of the tube of orbiting particles by the use of an exponential of the following form:
$\exp(-\gamma R\theta_{0,m}^\pm)$. Accordingly, the damping factor $a(k,\theta_s)$ can be split in the 
product of two terms as follows: $g(k)\,\exp(-\gamma R\theta_{0,m}^\pm)$. Coming back to the scattering
angle $\theta_s$, we rewrite expression (\ref{ventitre}) as follows (see also formulae (\ref{ventuno}) and
(\ref{ventidue})):
\beq
\label{ventiquattro}
f(k,\theta_s)=-\rmi g(k) \sum_{m=0}^\infty (-1)^m\,e^{\rmi 2\pi m\mu}\,
\frac{e^{\rmi\mu\theta_s}-\rmi e^{\rmi\mu(2\pi-\theta_s)}}{\sqrt{\sin\theta_s}}\,,
\eeq
$(0<\theta_s<\pi)$, where $\mu=R(k+\rmi\gamma)$. Next, we use the following expansion:
\beq
\label{venticinque}
\frac{1}{2\cos\pi\mu} = e^{\rmi\pi\mu}\sum_{m=0}^\infty (-1)^m\,e^{\rmi 2\pi m\mu}\,,~~~~~(\Imag\mu>0).
\eeq
By using equality (\ref{venticinque}) we can rewrite formula (\ref{ventiquattro}) as follows:
\beq
\label{ventisei}
f(k,\theta_s)=-g(k)\,e^{\rmi\pi/4}\,
\frac{e^{-\rmi[\mu(\pi-\theta_s)-\pi/4]} + e^{\rmi[\mu(\pi-\theta_s)-\pi/4]}}{2\cos\pi\mu\sqrt{\sin\theta_s}}\,,
\eeq
$(0<\theta_s<\pi)$. The r.h.s. of formula (\ref{ventisei}) contains the asymptotic behaviour of $\sqrt{2\pi\mu}$
times the Legendre functions $P_{\mu-1/2}(-\cos\theta_s)$, for $|\mu|\rightarrow\infty$ and
$|\mu|(\pi-\theta_s)>>1$ (see Ref. \cite{Erdelyi}). 
Then, writing $P_{\mu-1/2}(-\cos\theta_s)$ in place of its asymptotic behaviour,
we have for $|\mu|\rightarrow\infty$ and $0<\theta_s<\pi$:
\beq
\label{ventisette}
f(k,\theta_s)=-g(k)\,e^{\rmi\pi/4}\,
\frac{\sqrt{2\pi\mu} \, P_{\mu-1/2}(-\cos\theta_s)}{2\cos\pi\mu}\,.
\eeq
Finally, by putting: $\mu-1/2=\lambda$, we obtain, for $0<\theta_s\leq\pi$,
\beq
\label{ventotto}
f(k,\theta_s)=g(k)\,e^{\rmi\pi/4}\,\frac{\sqrt{\pi}}{2}\,\sqrt{2\lambda+1}\,
\frac{P_{\lambda}(-\cos\theta_s)}{\sin\pi\lambda}.
\eeq
It is then convenient to rewrite formula (\ref{ventotto}) in the following, more conventional, form:
\beq
\label{ventinove}
f(E,\theta)=G(E)\,
\frac{P_{\lambda(E)}(-\cos\theta)}{\sin\pi\lambda(E)}\,,~~~~~(0<\theta\leq\pi),
\eeq
where the scattering angle is now simply denoted by $\theta$, and 
$G(E)=\frac{\scriptstyle 1}{\scriptstyle 2}\sqrt{\pi}e^{\rmi\pi/4}g(k)\sqrt{2\lambda+1}$ is
the factor referred to the energy $E$.
The r.h.s. of formula (\ref{ventinove}) is singular at $\theta=0$, where $P_{\lambda(E)}(-\cos\theta)$ presents
a logarithmic singularity \cite{Sommerfeld}; therefore, representation (\ref{ventinove}) is only an approximation,
since a background compensating term that makes the amplitude finite and regular at every value of $\theta$ is
neglected. Nevertheless this approximation is very useful because it can describe a sequence of resonances and
bound states. In fact, if $\Real\lambda$ is integer and $\Imag\lambda<<1$, then the denominator $\sin\pi\lambda(E)$
is close to zero and, accordingly, we observe a bump in the cross--section: i.e., we have an orbiting resonance.
Since $\lambda$ is a function of the energy, representation (\ref{ventinove}) describes a sequence of resonances
which are obtained whenever $\Real\lambda$ crosses integer values, while  $\Imag\lambda$ remains very small. As 
we shall see in section \ref{phenomenological_section}, $\Imag\lambda$ tends to zero as $E$ tends to zero. 
For negative energy $\Imag\lambda=0$ and we have bound--states instead of resonances whenever $\Real\lambda$ 
crosses integer values. At the end, we obtain a pole trajectory connecting bound--states to resonances as it will 
be shown in section \ref{phenomenological_section}. 

\section{Orbiting Resonances and Bound--States}
\label{orbiting_section}
In this paper we focus our attention only on those effects that contribute to the more relevant component of the
spectrum: the orbiting resonances and the bound--states. However, they do not exhaust the entire phenomenology:
in particular, there are also the ``direct component'' of the collision which is
essentially composed by direct reflections, and the forward glory contribution. The former component does not
require particular explanations; instead, it is worth spending a few words about the forward glory, especially
in relation with the so--called ``glory undulations''. The latter effect
is produced by those paths that, moving through attractive and repulsive regions of the potential, undergo
equal positive and negative deflections and finally emerge at zero angle. Then, the interference between such paths
and the forward--diffracted trajectories with large impact parameter give rise to forward glory oscillations that
appear as undulations (glory undulations) in the total cross--section \cite{Nussenzveig1,Pauly}. 
The total number of oscillations can be related to the number of bound--states in the potential \cite{Pauly}. 
This relationship is one of the motivations of the great attention devoted to the ``glory--undulations'' by 
several authors \cite{Bernstein}.
In our analysis we do not consider the ``direct component'' of the scattering, which can be globally described
by a ``background--term'', which we neglect.

More generally, let us briefly recall that, in classical scattering theory, if the deflection function 
goes through zero (or through a negative multiple of $\pi$ for a non--zero value of the impact parameter) 
the differential cross--section diverges like $(\sin\Theta)^{-1}$ ($\Theta$ being the classical deflection angle).
The divergence can occur either in the forward direction $\Theta=-2n\pi$ ($n=0,1,2,\ldots\,$; forward glory), or 
in the backward
direction $\Theta=-(2n+1)\pi$ ($n=0,1,2,\ldots\,$; backward glory). Returning to the geometrical theory of the
orbiting outlined in the previous section, we see that, up to formula (\ref{ventisei}), $f(k,\theta_s)$
presents singularities at $\theta_s=0$ and $\theta_s=\pi$, which are due to the factor $(\sin\theta_s)^{-1/2}$.
Next, passing from formula (\ref{ventisei}) to representation (\ref{ventisette}) we introduce the Legendre
functions $P_\lambda(-\cos\theta_s)$ which are logarithmically singular at $\theta_s=0$, but regular
at $\theta_s=\pi$. The smoothing of these singularities corresponds to the passage from the
classical to the semi--classical approximation. As we have already noted several times, the obtained representation
is far from being complete, as it does not represent the direct components of the collision:
diffraction, direct reflection and forward glory contribution \cite{Nussenzveig1}.

Hereafter we will show that representation (\ref{ventinove}), in spite of the limits mentioned above,
can, nevertheless, represent an ordered sequence of orbiting resonances. With this in mind, we project
the amplitude (\ref{ventinove}) on the $l$-th partial wave, obtaining:
\beq
\label{3-uno}
a_l=\frac{e^{2\rmi\delta_l}-1}{2\rmi k}=\frac{G(E)}{\pi}\,
\frac{1}{(\alpha+\rmi\beta-l)(\alpha+\rmi\beta+l+1)}\,,
\eeq
where $\alpha+\rmi\beta=\lambda$. Next, when the elastic unitarity condition may be applied, we get the following
relationship among $G$, $\alpha$ and $\beta$:
\beq
\label{3-due}
G(E) = -\frac{\pi}{k}\,\beta\,(2\alpha+1)\,.
\eeq
From (\ref{3-uno}) and (\ref{3-due}) we obtain
\beq
\label{3-tre}
\delta_l=\sin^{-1}\,
\frac{\beta(2\alpha+1)}{\{[(l-\alpha)^2+\beta^2][(l+\alpha+1)^2+\beta^2]\}^{1/2}}\,.
\eeq
Formula (\ref{3-tre}) represents a sequence of orbiting resonances: i.e., an ordered set of phase--shifts that
cross $\delta_l=\pi/2$ with positive derivative.

The bound--states are not directly revealed by scattering experiments. We know, however, from the generalized
Levinson theorem that $\delta_l(0)=n_l\pi$, where $n_l$ is the number of bound--states with angular momentum $l$,
including those states which are not admitted by the Pauli principle \cite{Swan}. 
Hereafter we shall consider, however, 
atom--atom scattering where the Pauli excluded states are not involved. Furthermore, in the cases considered
in the next section we shall deal with a sequence of phase--shifts whose zero--energy value is exactly
$\pi$. This implies that there is a sequence of bound--states with increasing value of the angular momentum
$l$. We shall see that these bound--states can be connected with continuity to the
sequence of orbiting resonances, obtaining thus a band of bound--states and resonances (rotational band),
generated by a unique pole moving in the complex angular momentum plane. At this point we are faced with the
problem of finding a representation which is able to describe either those phase--shifts that do not 
resonate but decrease from their zero--energy value, that is $\pi$, and cross $\pi/2$ downward,
and also that part of the phase--shifts that decrease after a resonance.
Since these phase--shifts are due to the
repulsive component of the potential, we can assume, in a very rough model, that they are produced by a
hard--core potential. In the subsequent section we shall see that this model, in spite of its roughness, is in
reasonable agreement with the phenomenological data.

First, let us recall that the phase--shifts produced by an impenetrable sphere of radius $r_0$ are given 
by \cite{Mott}
\beq
\label{3-quattro}
\delta_l = \tan^{-1}\,\frac{J_{l+1/2}(kr_0)}{N_{l+1/2}(kr_0)}\,,
\eeq
where $J_{l+1/2}$ and $N_{l+1/2}$ denote the Bessel and Neumann functions, respectively. It is
easy to check that when $kr_0=(l+1)\pi/2$, then the phase--shift $\delta_l$, given by formula (\ref{3-quattro}),
equals $(-\pi/2)$; but the phenomenological data (see the next section) indicate that the radius $r_0$ of the 
hard--core, which models the repulsive component of the interaction, varies with $l$. 
We can then write $kr_l$ instead of $kr_0$, and relate this expression
to the angular momentum $[\alpha(\alpha+1)]^{1/2}$ through the equality $kr_l=B_l[\alpha(\alpha+1)]^{1/2}$,
where $B_l$ is a phenomenological parameter. 
We then glue together the term describing the orbiting to the hard--core phase--shifts as follows:
\beq
\label{3-cinque}
\delta_l=\sin^{-1}\,\frac{\beta(2\alpha+1)}{\{[(l-\alpha)^2+\beta^2][(l+\alpha+1)^2+\beta^2]\}^{1/2}}
+A_l\tan^{-1}\,\frac{J_{l+1/2}(B_l[\alpha(\alpha+1)]^{1/2})}{N_{l+1/2}(B_l[\alpha(\alpha+1)]^{1/2})}\,,
\eeq
where $A_l$, as well as $B_l$, depends on $l$, and both can be regarded as fitting parameters. 
As we shall show in the next section, formula (\ref{3-cinque}) is able to reproduce an ordered sequence of orbiting
resonances, and also gives in the correct order the downward crossing of the phase--shifts $\delta_l$ 
through $\pi/2$.
The parameter $A_l$ in formula (\ref{3-cinque}) accounts for the strength of the repulsive action which 
models the singularity of the potential at the origin. This strength decreases as the angular momentum increases, 
in agreement with the fact that the radius of the hard--core increases with $l$. 
From the phenomenological analysis the dependence of $A_l$ and $B_l$ on $l$ turns out to be nearly linear 
(see figs. \ref{f3}(c) and (d)).

It is interesting to notice that a similar approach has been applied also in nuclear physics, in 
particular in the $\alpha$--$\alpha$ and $\alpha$--$^{40}$Ca elastic scattering (see refs. \cite{Viano,Fioravanti}).
In these works the hard--core--type interaction has been used in place of the more realistic, but less tractable,
exchange--forces which, ultimately, give rise to non--local potentials. However, in this case the approximation
is not very faithful; in fact, while the fits of the rotational resonances are very satisfactory, for
what concerns the downward passage of the phase--shifts, the theory yields only the general trend, but lacks for
details. It is currently in progress an analysis of the complex angular momentum singularities
associated with non--local potentials which aims at obtaining a more accurate representation of the downward
behaviour of the nuclear phase--shifts.

\section{Phenomenological Analysis}
\label{phenomenological_section}
In this section a series of numerical fits, performed by using formula (\ref{3-cinque}), will be presented. 
The experimental data being considered concern the H--Kr system, and are taken from Ref. \cite{Toennies};
in particular, we refer to figure 3 of that reference.
In formula (\ref{3-cinque}) the quantities $A_l$ and $B_l$ are considered as free fitting parameters, and
for what concerns the functions $\alpha(E)$ and $\beta(E)$ we adopt the following parametrizations:
\begin{eqnarray}
\label{4-uno}
&&\alpha(\alpha+1) = 2IE+\alpha_0\,, \\
\label{4-due}
&&\beta = d_1 \alpha(E)\sqrt{E} + d_2\sqrt{E}\,,
\end{eqnarray}
where $I=\mu R^2$ is the moment of inertia, $\mu$ is the reduced mass, $R$ the interparticle distance, and $E$ is
the center of mass energy; the parameters $d_1$ and $d_2$, that from the analysis turn out to be nearly independent 
of $l$, and the intercept $\alpha_0$ are to be determined through the fits of the experimental phase--shifts
(see the figure legends for numerical details).

\begin{figure}[ht]
\centering
\includegraphics[width=3.4in]{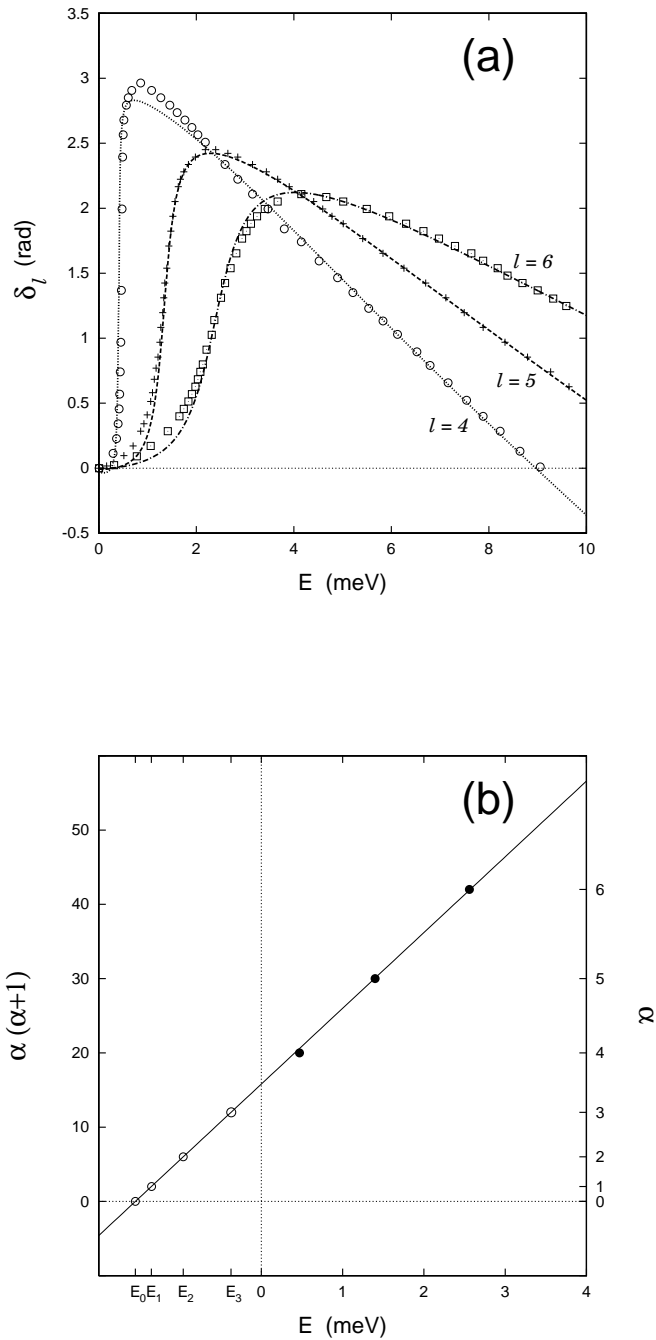}
\caption{
H-Kr system. (a): Observed phase--shifts \cite{Toennies} and corresponding fits 
(see formula (\ref{3-cinque})) for the partial waves $l=4,5,6$ versus the energy 
of the center of mass $E$.
$d_1=0.13$ (meV)$^{-1/2}$; $d_2=-0.48$ (meV)$^{-1/2}$ (see formula (\ref{4-due})).
$l=4$: observed ($\circ$), fitted (dotted line).
$l=5$: observed ($+$), fitted (dashed line).
$l=6$: observed ($\square$), fitted (dash--dotted line).
(b): $\alpha(\alpha+1)$ vs. $E$ (see formula (\ref{4-uno})). 
$I=5.1$ (meV)$^{-1}$; $\alpha_0=15.8$.
The filled dots indicate the energy $E_l$ of the resonance for the partial waves $l=4,5,6$.
The open circles indicate the energy values $E_l$ of the bound states for $l=0,1,2,3$
extrapolated from the line $\alpha(\alpha+1)$:
$E_0=-1.55$ meV, $E_1=-1.35$ meV, $E_2=-0.96$ meV, $E_3=-0.37$ meV.
}
\label{f2}
\end{figure}

In figure \ref{f2}(a) the fits of the phase--shifts for $l=4,5,6$, which display orbiting resonances, are shown.
The upward crossings of $\delta_l=\pi/2$, i.e. the energies of the resonances, are reported in figure \ref{f2}(b)
in correspondence of the integer values of $\alpha=4,5,6$ (filled dots), along with their interpolation
with the line $\alpha(\alpha+1)$ as a function of $E$. The good agreement between resonance locations ($E>0$)
and the line $\alpha(\alpha+1)$ strongly supports the correctness of the linear parametrization (\ref{4-uno}).
Since $\alpha_0>0$, we can extrapolate, for $E<0$, the location of the bound--states corresponding to $l=0,1,2,3$
(open circles).

\begin{figure}[ht]
\centering
\includegraphics[width=6in]{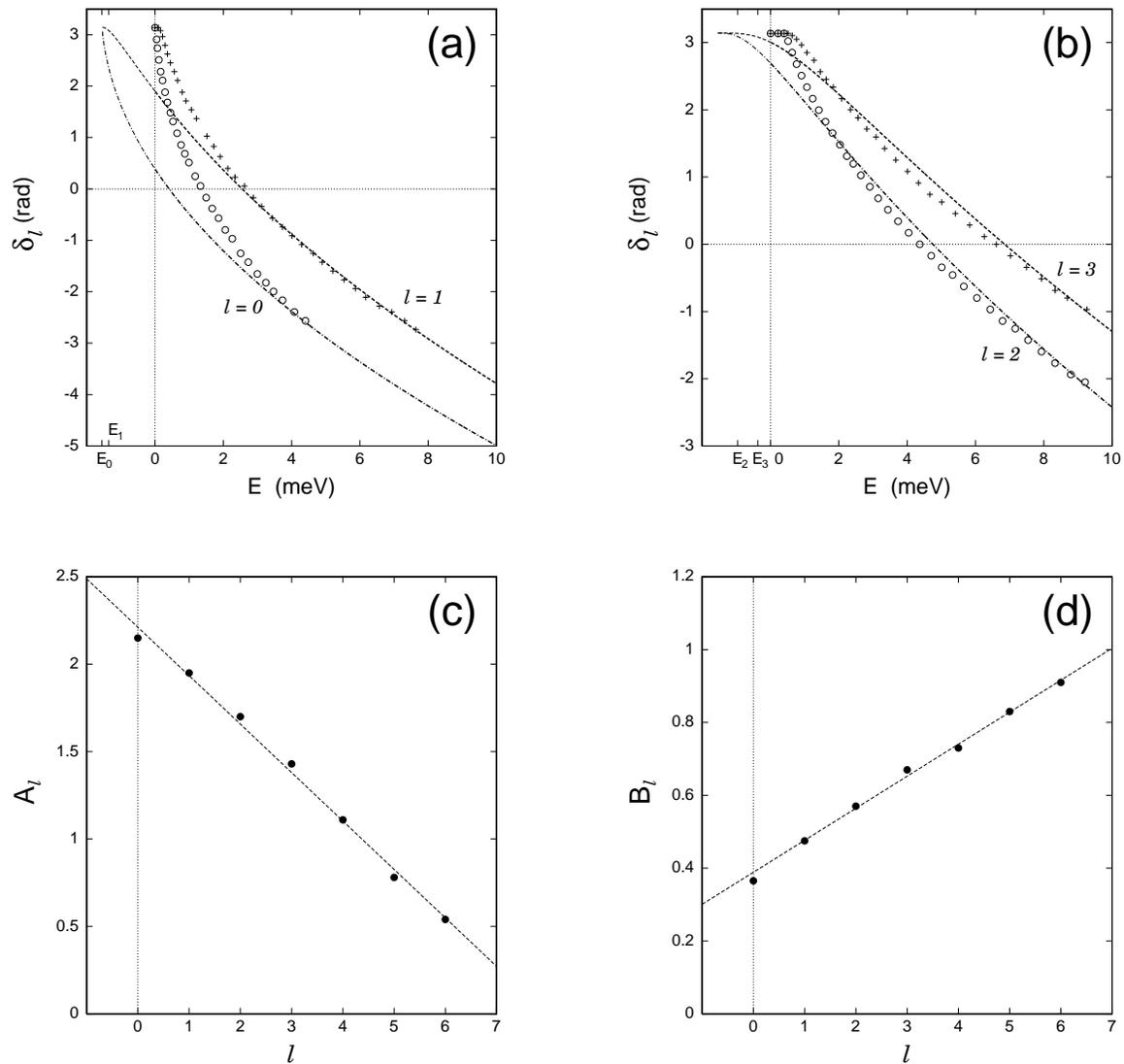}
\caption{
H-Kr system. 
Observed phase--shifts \cite{Toennies}
and corresponding fits (see formula (\ref{3-cinque}))
for the partial waves $l=0,1,2,3$ versus the energy of the center of mass $E$.
For what concerns the meaning of $E_0$, $E_1$, $E_2$ and $E_3$ see the legend of figure \ref{f2}.
The values of the fits for $E<0$ (thinner lines) must be understood as explained in section
\ref{phenomenological_section}.
(a): Case $l=0,1$.
$l=0$: observed ($\circ$), fitted (dash--dotted line).
$l=1$: observed ($+$), fitted (dashed line).
(b) Case $l=2,3$.
$l=2$: observed ($\circ$), fitted (dash--dotted line).
$l=3$:  observed ($+$), fitted (dashed line).
(c): $A_l$ vs. $l$. 
(d): $B_l$ vs. $l$. See formula (\ref{3-cinque}).
}
\label{f3}
\end{figure}

In figures \ref{f3}(a) and (b) the fits of the phase--shifts for $l=0,1$ and $l=2,3$ are shown. The high energy 
behaviour is well reproduced, whereas the Levinson value at $E=0$ (i.e., $\delta_l=\pi$), which is clearly 
satisfied by the experimental data, is not reached by the fitting curves. 
On the other hand, it is quite tempting to notice that if we extrapolate
these theoretical phase--shifts to the unphysical region at negative energy (in spite of the fact that
the phase--shifts have physical meaning exclusively at non--negative energy), they get close to the value
$\delta_l=\pi$ approximately at the energies $E_0$, $E_1$, $E_2$ and $E_3$, that closely correspond to the 
location of the bound--states that have been recovered from figure \ref{f2}(b), and whose angular momenta are 
precisely $l=0,1,2,3$.
In figures \ref{f3}(c) and (d) the values of the parameter $A_l$ and $B_l$ that result from the fits of 
figures \ref{f2}(a), \ref{f3}(a) and \ref{f3}(b) are displayed as functions of $l$. 
$A_l$ presents a nearly linear decrease with $l$, while $B_l$ increases almost linearly with $l$. 

Concluding, we illustrate the relationship between the function $\beta$ and the widths $\Gamma$ of the 
orbiting resonances. These quantities are related to the tunnelling across the centrifugal barrier, 
which is now a complex quantity. 
We have
\beq
\label{4-tre}
\Imag \left<\frac{\lambda(\lambda+1)}{2\mu R^2}\right>=\beta(2\alpha+1)\,\frac{1}{<2\mu R^2>}=\frac{\Gamma}{2}\,,
\eeq
from which it follows
\beq
\label{4-quattro}
\Gamma = \frac{\beta(2\alpha+1)}{I}\,,~~~~~~(I= \mbox{moment of inertia}).
\eeq
From (\ref{4-quattro}) $\Gamma$ can be easily evaluated, since $\alpha(E)$, $\beta(E)$, and $I$ are recovered 
from the fits of the phase--shifts; the values of $\Gamma$ are respectively: 
$l=4$, $\Gamma=4.69 \cdot 10^{-2} \, {\rm meV}$;
$l=5$, $\Gamma=4.45 \cdot 10^{-1} \, {\rm meV}$;
$l=6$, $\Gamma=1.30 \, {\rm meV}$.

\section{Final Remarks}
\label{remarks_section}
In conclusion, we want to stress the difference between orbiting and diffracted rays. 
The diffracted rays can be explained as follows: when a ray grazes a boundary surface, it splits in two: 
one part keeps going as an ordinary ray, whereas the other part travels along the surface. 
At every point along its path this ray splits in two again:
one part proceeds along the surface, and the other one leaves the surface along the tangent to the surface
itself. This splitting can be mathematically explained as due to the non--uniqueness of the Cauchy problem for
geodesics in a Riemannian manifold with boundary (see Ref. \cite{DeMicheli}). 
The damping factors can then be related to the curvature of the diffracting body \cite{DeMicheli}. 
Conversely, for what concerns the orbiting case, the attenuation of the current of orbiting particles
is produced by the leakage of the ray tube of the particles which describe circular orbits, and which is due to the 
tunnelling across the centrifugal barrier. It follows that the nature and the extent of the damping factors are
drastically different in these two phenomena. We can say that, while the diffracted rays are produced by edge
effects, this is not at all the case for the orbiting phenomena. 
Edge effects are indeed relevant in optics where the diffracting body (i.e., the obstacle) is compact, but
in particle scattering they can be observed only in those processes described by short--range potentials,
e.g. decreasing exponentially in $r$, which can simulate boundary effects.
This does not happen in molecular scattering where the potentials that describe properly the phenomena
are of Lennard--Jones type, which decrease, for large values of $r$, only as an inverse power.
Thus, we can say that the near--forward diffractive scattering by a potential of this type, that is
having a long--range tail, is not due to edge diffraction, but is rather produced by the contributions of
very large angular momentum paths that undergo small deflection caused by the tail of the potential
(see Ref. \cite{Nussenzveig1}).

Finally, it is worth remarking that in the orbiting phenomena and in the case of diffracted rays, 
the attenuation can be explained merely as a damping of the amplitude, while the phase
remains real: in both cases we can properly speak of real rays. However, one is tempted to explain the 
attenuation of the intensity of the flux by introducing a complex phase, and, accordingly complex rays. 
We believe that this approach is not at all convenient for describing orbiting and diffraction phenomena, 
where the attenuation of the amplitude can be evaluated through the transport equation. 
Conversely, complex rays emerge, as appropriate mathematical solutions of the eikonal equation, for representing 
the shadow of the caustic. Therefore, they play a relevant role in total reflection, where they describe the 
exponentially damped penetration into the rarer medium associated to the surface waves traveling along the boundary. 
In molecular scattering they are again present in rainbow phenomena, and precisely in the shadow of the rainbow caustic. 
In fact, we can show that in the neighborhood of a caustic the nature of the eikonal equation differs from the 
standard one: its characteristics, which are real in the illuminated region, become complex conjugate 
beyond the caustic (i.e., in the shadow of the caustic), and correspond to the complex rays.

\newpage

\end{document}